\documentclass[manuscript]{acmart}

\usepackage[utf8]{inputenc} 
\usepackage[T1]{fontenc}    
\usepackage{hyperref}       
\usepackage{url}            
\usepackage{booktabs}       
\usepackage{amsfonts}       
\usepackage{nicefrac}       
\usepackage{microtype}      
\usepackage{xcolor}         
\usepackage{multirow}
\usepackage{graphicx}
\usepackage{tabularx}
\usepackage{subcaption}

\usepackage[compact]{titlesec}
\titlespacing{\subsection}{0pt}{1.1ex}{0ex}

\title{Multi-Task Learning For Reduced Popularity Bias In Multi-Territory Video Recommendations}

\author{Phanideep Gampa}
\authornote{Both authors contributed equally to this research.\\ Copyright held by the author(s).}
\email{phanide@amazon.com}
\author{Farnoosh Javadi}
\authornotemark[1]
\email{javadif@amazon.com}
\affiliation{%
  \institution{Amazon Prime Video}
  \country{United States}
}

\author{Belhassen Bayar}
\affiliation{%
  \institution{Amazon Prime Video}
    \country{United States}
  }
\email{bayarb@amazon.com}

\author{Ainur Yessenalina}
\affiliation{%
  \institution{Amazon Prime Video}
  \country{United States}
  }
\email{yessenal@amazon.com}

\begin{document}

\begin{abstract}
Various data imbalances that naturally arise in a multi-territory personalized recommender system can lead to a significant item bias for globally prevalent items. A locally popular item can be overshadowed by a globally prevalent item. Moreover, users' viewership patterns/statistics can drastically change from one geographic location to another which may suggest to learn specific user embeddings. In this paper, we propose a multi-task learning (MTL) technique, along with an adaptive upsampling method to reduce popularity bias in multi-territory recommendations. Our proposed framework is designed to enrich training examples with active users representation through upsampling, and capable of learning geographic-based user embeddings by leveraging MTL.
Through experiments, we demonstrate the effectiveness of our framework in multiple territories compared to a baseline not incorporating our proposed techniques.~Noticeably, we show improved relative gain of up to $65.27\%$ in PR-AUC metric. A case study is presented to demonstrate the advantages of our methods in attenuating the popularity bias of global items.
\end{abstract}

\ccsdesc[500]{Computing methodologies~Transfer learning}
\ccsdesc[300]{Computing methodologies~Supervised learning by classification}

\keywords{neural networks, video recommendations, two-tower networks, multi-task learning}
	
	\maketitle

\section{Introduction}
In modern streaming services such as YouTube, Prime Video, Netflix and etc., recommender systems (RS) have become the cornerstone of user engagement by helping the users find relevant contents~\cite{amatriain2015recommender,berkovsky2017recommend}. ML-based RS are typically trained in a supervised way using the logs of previous user-item interactions as explicit or implicit feedback of user’s interest ~\cite{gomez2015netflix,cheng2016wide}. For RS with wide geographical coverage across multiple countries such as the Amazon Prime Video, theoretically-speaking, one can separately train multiple region-specific ML models to perform recommendations for different regions. Maintaining multiple separate models is challenging from the ML operational perspective. It is therefore, highly preferable for a large-scale recommendation system to create one global model that can learn to make useful recommendations across different geographical regions.

One common challenge while building ML-based RS is the various imbalances that naturally occur in the user-item interaction training data~\cite{chen2020bias,gomez2020characterizing}. In this work, we are particularly concerned with the data imbalances that exist across different geographical partitions of a global RS with worldwide coverage. This data imbalance can occur in any multi-territory setting. For instance, one video title can be the 10’th popular title in one territory but still can have much more user interaction history than the most popular video title in other territories. Also, user interaction history data of a video title that is only streamed in one specific territory (i.e., a local item) can be significantly smaller than that of a video title that is streamed in multiple territories (i.e., a global item), even though the local item/title is much more popular in that specific territory. If these imbalances are not properly handled, the global RS will inevitably suppress the locally popular items at the expense of globally present items. In other words, a single global model would suffer from popularity biases towards globally present items. 

Popularity bias arising from the data imbalance decreases the fairness of the recommendations~\cite{abdollahpouri2019unfairness,abdollahpouri2020multi}, specifically for the titles local to a certain country. And the bias would only amplify in a closed feedback setting~\cite{chen2020bias,zhu2021popularity}, where the titles present in a lot of territories are always ranked higher and the gap between the global and local titles would either remain the same or increase. To prevent the model from overfitting on the global titles and to improve the fairness of local titles, we propose the use of Hard Parameter Sharing based Multi Task Learning (MTL)~\cite{caruana1993multitask} approach. MTL has strong regularisation properties~\cite{ruder2017overview,baxter1997bayesian} that improves the generalisation power of the model by preventing overfitting on global titles. At the same time, MTL learns shared representations that can be used by countries with less data allowing the transfer of knowledge from countries with large data. 

Rebalancing and regularisation are among the frequent techniques used for solving the unfairness issue arising from the data imbalance~\cite{chen2020bias} in the field of recommendation. We address the popularity bias issue in our system by means of a data-centric and a model-centric approach, which are our key contributions, as follows: (1) A data upsampling strategy based on active users criteria to enrich the training data with more number of informative examples to learn from. (2) An MTL approach to learn the refined user representations at a territory group level to deal with the distributional differences of viewership patterns across different territories. (3) Through a case study, we show how MTL reduces the popularity bias of global titles by learning group-specific parameters from the conditional distributions at a territory group level.



Our baseline model architecture consists of a two-tower Neural Network (NN) model ~\cite{huang2013learning} (one tower for user, and one for title), trained to learn propensity score for a user’s first time content discovery by leveraging representation of user and title embeddings as model input; which are vended by an upstream representation learning feature store. The exact details are described in section~\ref{subsec:input_embeddings}.
Through a set of experiments, we compared our baseline recommendation system to a model that incorporates the MTL approach and upsampling strategy. Our experimental results demonstrate the advantage of our proposed model which significantly improved the PR-AUC for the majority of territories, and also helped to reduce the popularity bias towards global titles.
\label{section:introduction}

\section{Related Work}



\textbf{Popularity Bias.}~Popularity bias in ML-based recommender systems occurs when popular items are typically recommended even more frequently than their organic popularity level~\cite{abdollahpouri2020multi,abdollahpouri2019unfairness}.
Various methods have been proposed to mitigate the effects of popularity bias such as  Regularization~\cite{abdollahpouri2017controlling,chen2020esam,kamishima2014correcting}, Adversarial Learning~\cite{krishnan2018adversarial} and Causal Graphs~\cite{wang2021deconfounded,wei2021model}.

\textbf{Unfairness.}~A topic that is closely related to the popularity bias is the fairness of ML-based systems. Many ML-based systems have been shown to inadvertently discriminate against certain individuals or groups due to the intrinsic bias that exists in the training data~\cite{friedman1996bias,lin2019crank}. Various works have proposed mitigating the ML-induced unfairness by means of techniques such as   Rebalancing~\cite{asudeh2019designing,biega2018equity,geyik2019fairness}, Regularization~\cite{kamishima2012enhancement,kamishima2013efficiency,kamishima2016model},
Adversarial Learning~\cite{beigi2020privacy,bose2019compositional,edwards2015censoring} and 
Causal Modeling~\cite{nabi2018fair,wu2018discrimination,wu2019counterfactual}.

\textbf{Multi-Task Learning.}~Multi-task learning (MTL) has been utilized in many Machine Learning applications such as natural language processing and speech recognition~\cite{collobert2008unified,deng2013new}. In the related works, MTL has been primarily shown to provide benefits such as improved regularisation capabilities along with efficient transfer learning~\cite{weiss2016survey,torrey2010transfer}.
There are two main MTL model architectures in Deep neural networks: hard parameter sharing and soft parameter sharing. In hard parameter sharing, a number of hidden layers are shared among all tasks. These common layers are followed by a set of task-specific layers~\cite{long2017learning,hashimoto2016joint}. In soft parameter sharing, on the other hand, each task has a separate model with separate parameters and a regularization term is used in order to encourage the parameters of different models to be close~\cite{misra2016cross}. In this work, we selected the hard parameter sharing architecture because of its regularization capabilities and its advantages of refined task representation learning from a shared part. To the best of our knowledge, our work is the first application of MTL to mitigate the item popularity bias in a global RS.

\begin{figure}[htbp]
    \centering
        \includegraphics[clip, trim=5cm 17cm 6cm 3cm, scale=0.90]{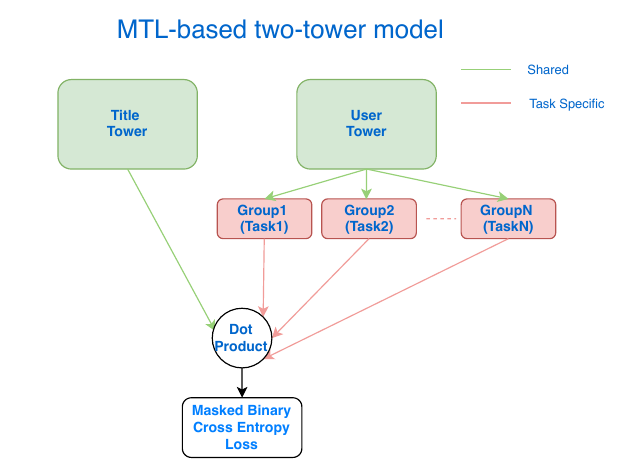}
    \caption{Block diagram of the proposed MTL-based two-tower model where the user tower is branched out based on the territory grouping. Shared neural network blocks are in green and task-specific blocks are in red. Given a random batch, the loss function is a masked summation of binary cross entropy loss for each task.}
    \label{fig:MTL Block}
\end{figure}

\section{Approach}
\subsection{Input Embeddings}
\label{subsec:input_embeddings}
For all our experiments, we use user and title vector embeddings extracted from other pre-trained models. For the title specific data, we first extract 768 dimensional vectors using a fine-tuned BERT~\cite{devlin2018bert} applied on each title's textual metadata such as name, genres, and synopsis. Next, several content-related categorical vectors are appended to each title's BERT representation. For users, we use 512-dimensional latent vectors extracted from a ResNet architecture as described in~\cite{wang2021exploring}. This model was trained to encode contextual data of Prime Video users, including long-term and near-term streaming history, territory and other features.
\subsection{Data Upsampling using Active Users}
\label{subsec:Upsampling}
The data creation process generally comprises of selecting the set of users and a set of titles, and then generating the positive and negative user-title pairs. A random sampling of users from the customer universe across all the territories would be biased towards the territories with large number of users. At the same time, the set of active users can have minor representation than the non active streamers. Active users are those who watched at least one title in the recent few months. To reduce these disparities, we add a disjoint set of active users sampled independently to the random user set. This would enrich our training dataset with more number of active user embeddings. Also, this would improve the likelihood of having more number of positive examples in training particularly helping the minor territories.


\subsection{MTL-based Two-Tower Model}
\label{subsec:MTL two tower model}
To mitigate the issues arising from data imbalance and to better deal with the distributional differences across territories, we propose the use of MTL-based two-tower model. We use a hard parameter sharing~\cite{caruana1993multitask} based architecture, where we branch out of the user tower to have task-specific layers, one for each territory group. For this approach, we first group the available 200+ territories into four groups based on the geo-partition they are present in. This criteria groups geographically closer territories together that can potentially have similar viewership patterns. At the same time, we also try to keep a good balance of major and minor territories in each group for efficient transfer learning. We refer a territory as major or minor based on its viewership volume. 

We propose a neural network architecture that consists of two parts, shared blocks across all territories and a task-specific block for each territory group as depicted in the figure~\ref{fig:MTL Block}. Specifically, shared blocks are the user tower and title tower as that of the baseline model. We then add separate task/group-specific neural network blocks on top of the user tower. During training time, the shared blocks are always updated using back propagation for all the data points. The group-specific blocks are only updated for the data points coming from the respective group. Such MTL-based techniques are known for having the advantages of transfer learning — minor territories learning from the major territories and vice versa —  happening through the shared part. At the same time, we also have independent blocks to take care of the distributional differences across the country groups. These task-specific blocks convert the shared user representations to the group-specific representation. This allows us to acknowledge the differences in the user viewership patterns across various territories and to localize the recommendations at a group level.

\begin{table}[h]
\caption{Statistics of number of positive data points in some major territories signifying the imbalance of positive rewards.}
\label{table: data}
\centering
\begin{tabular}{p{1.9cm} p{0.9cm} p{0.9cm} p{0.9cm} p{0.9cm} p{0.9cm} p{0.9cm} p{0.9cm} p{0.9cm}}
\toprule
 & T1 & T2 & T3 & T4 & T5 & T6 & T7& T8 \\
\midrule
\multirow{1}{*}{\shortstack[l]{Order of}} & 1M&1M&1M&1M&1M&1M& 100k&100k\\
\bottomrule

\toprule
& T9 & T10 & T11 & T12 & T13 & T14 & T15 & T16 \\
\midrule
\multirow{1}{*}{\shortstack[l]{Order of}} & 100k& 10k&10k&10k&100k&100k&100k&100k\\
\bottomrule
\end{tabular}
\end{table}

 \begin{table}[h]
\caption{Relative gain of PR-AUC score in various territories $T1,T2, \ldots , T15, T16$ for the MTL model evaluated on three different test sets. Positive average results are in bold for each territory.}
\label{table: results}
\begin{tabular}{p{1.9cm}  p{1.2cm}   p{0.85cm} p{0.85cm} p{0.91cm} p{0.85cm} p{0.85cm} p{0.92cm} p{0.85cm} p{0.85cm}}
\toprule
 & Test Set & T1 & T2 & T3 & T4 & T5 & T6 & T7& T8 \\
\midrule
\multirow{4}{*}{\shortstack[l]{MTL vs \\ Baseline}}&Test-1 & 9.88&34.05&-7.72&11.36&15.04&27.12&10.10&21.36\\
&Test-2 & 35.30&65.27& -21.18&24.58&24.73&23.22&0.73&-0.35\\
&Test-3 & 31.99&16.36&11.39&18.98&9.49&20.69&33.67&5.35\\
 \cmidrule{2-10}
&Average& \textbf{25.72}&\textbf{38.56}&-5.83&\textbf{18.30}&\textbf{16.42}&\textbf{23.67}&\textbf{14.83}&\textbf{8.78} \\
\midrule
\multirow{4}{*}{\shortstack[l]{MTL vs \\ Baseline  with \\  upsampling}}&Test-1 &7.27&23.46&-30.16&-3.98&4.62&-16.52&-0.65&6.30 \\
&Test-2 &9.80&20.36&-5.11&7.59&11.78&5.05&2.95&3.79\\
&Test-3 & 18.31&8.17&0.84&6.89&9.92&6.69&13.17&-3.23\\
 \cmidrule{2-10}
&Average& \textbf{11.79}&\textbf{17.33}&-11.48&\textbf{3.49}&\textbf{8.77}&-1.58&\textbf{5.15}&\textbf{2.28} \\
\bottomrule

\toprule
 & Test Set& T9 & T10 & T11 & T12 & T13 & T14 & T15 & T16 \\
\midrule
\multirow{4}{*}{\shortstack[l]{MTL vs \\ Baseline}}&Test-1 & 5.21&13.43&7.68&54.75&10.45&7.87&5.51&7.36\\\
&Test-2 & -5.85&46.97&12.88&21.64&36.42&32.31&20.20&25.68\\
&Test-3 & 3.75&-2.92&38.39&45.69&23.06&12.61&20.40&12.70\\
 \cmidrule{2-10}
 &Average& \textbf{1.03}&\textbf{19.16}&\textbf{19.65}&\textbf{40.69}&\textbf{23.31}&\textbf{17.59}&\textbf{15.37}&\textbf{15.24} \\
\midrule
\multirow{4}{*}{\shortstack[l]{MTL vs \\ Baseline  with \\  upsampling}}&Test-1 & 3.91&15.52&28.45&77.35&3.38&2.39&2.41&0.48\\
&Test-2 & 9.47&69.10&16.07&15.89&10.28&18.59&3.86&0.13\\
&Test-3 & -6.86&0.25&22.26&10.85&13.68&6.79&20.35&13.43\\
 \cmidrule{2-10}
&Average& \textbf{2.17}&\textbf{28.29}&\textbf{22.26}&\textbf{34.70}&\textbf{9.11}&\textbf{9.25}&\textbf{8.88}&\textbf{4.68} \\
\bottomrule
\end{tabular}
\end{table}

\begin{table}[h]
\caption{Title level PR-AUC gain for a set of local and global titles.}
\label{table: case study}
\centering
\begin{tabular}{ccc}
\toprule
 Title  & MTL vs Baseline & MTL vs Baseline Upsampled \\
\midrule
Local Title 1&6.50& 2.63\\
Local Title 2&3.83&6.23\\
Global Title 1&4.28& 3.15\\
Global Title 2&36.22& 35.27\\
\bottomrule
\end{tabular}
\end{table}

\begin{figure*}[h]
    
    \begin{subfigure}{0.495\textwidth}
    \centering
    \includegraphics[clip, trim=1cm 0.75cm 0.75cm 0.75cm, width=1.00\textwidth]{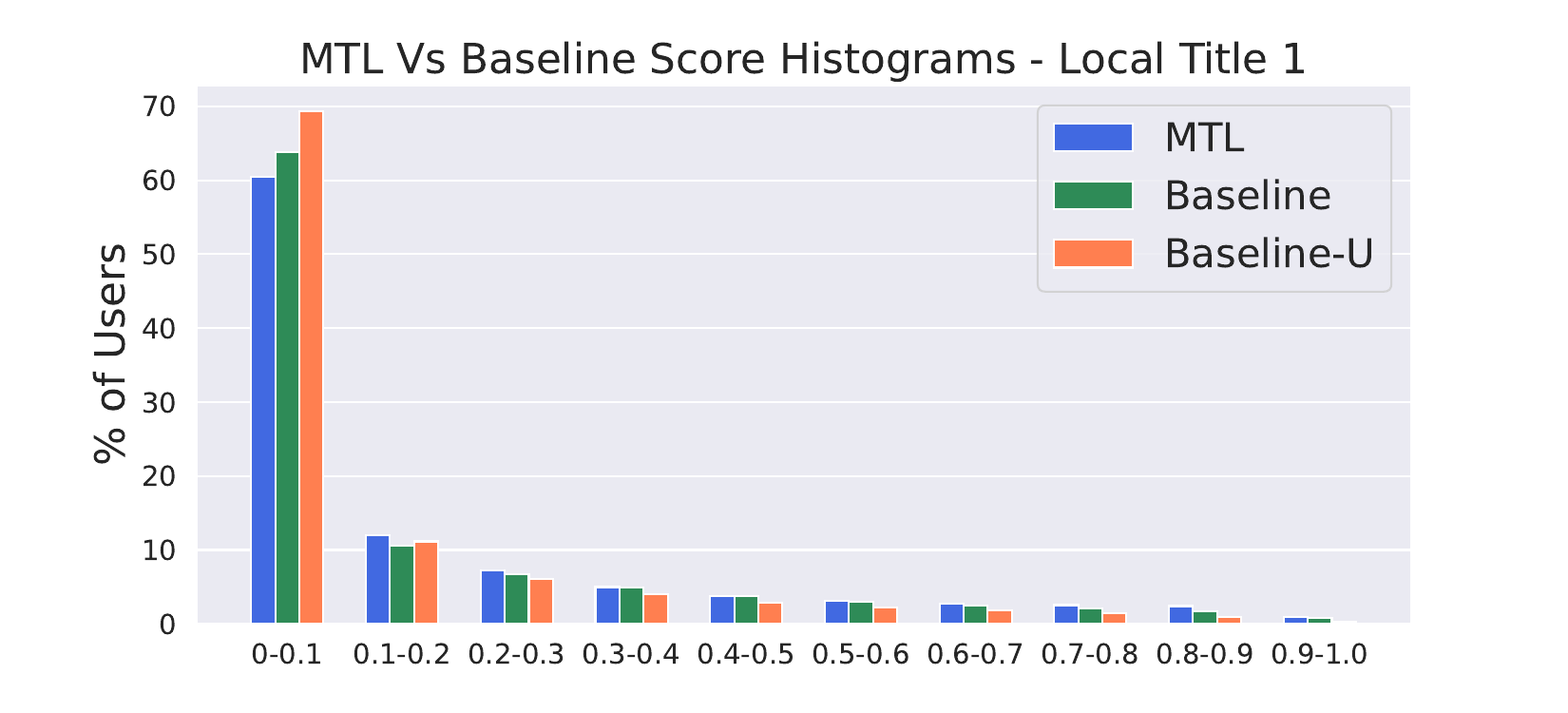}
    \caption{Local Title 1}
    \label{fig:es_local}
    \end{subfigure}
    \begin{subfigure}{0.495\textwidth}
    \centering
    \includegraphics[clip, trim=1cm 0.75cm 0.75cm 0.75cm, width=1.00\textwidth]{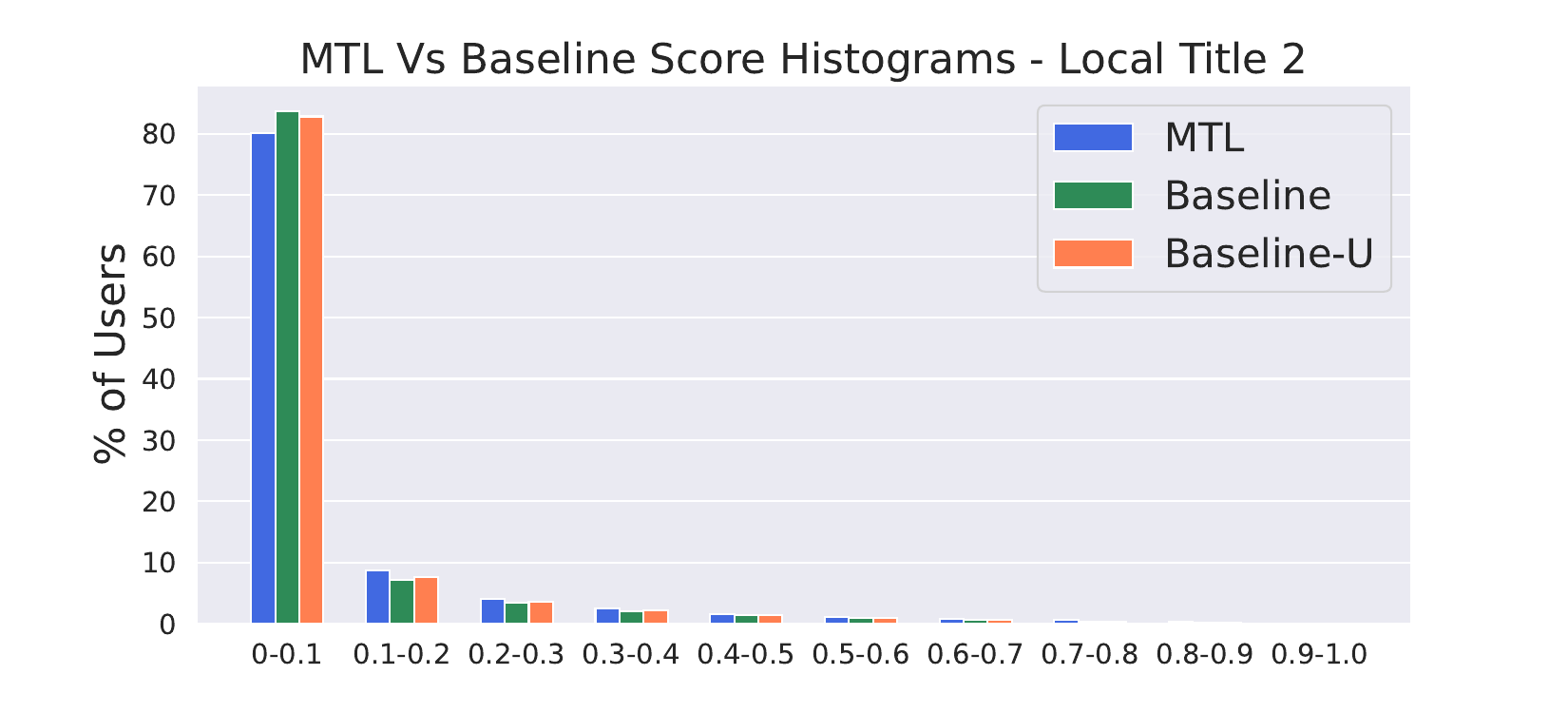}
    \caption{Local Title 2}
    \label{fig:br_local}
    \end{subfigure}
    \begin{subfigure}{0.495\textwidth}
    \centering
    \includegraphics[clip, trim=1cm 0.75cm 0.75cm 0.75cm, width=1.00\textwidth]{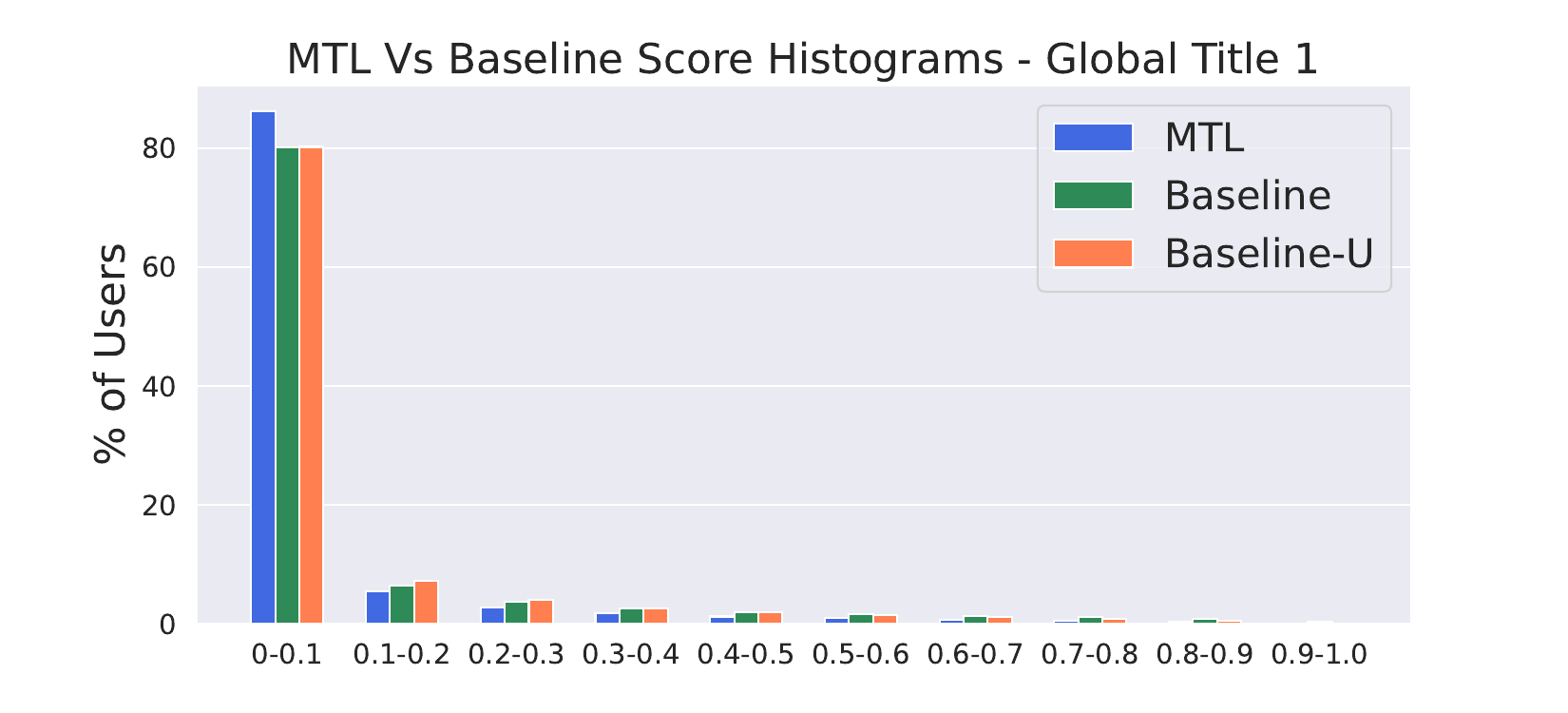}
    \caption{Global Title 1}
    \label{fig:es_global}
    \end{subfigure}
    \begin{subfigure}{0.495\textwidth}
    \centering
    \includegraphics[clip, trim=1cm 0.75cm 0.75cm 0.75cm, width=1.00\textwidth]{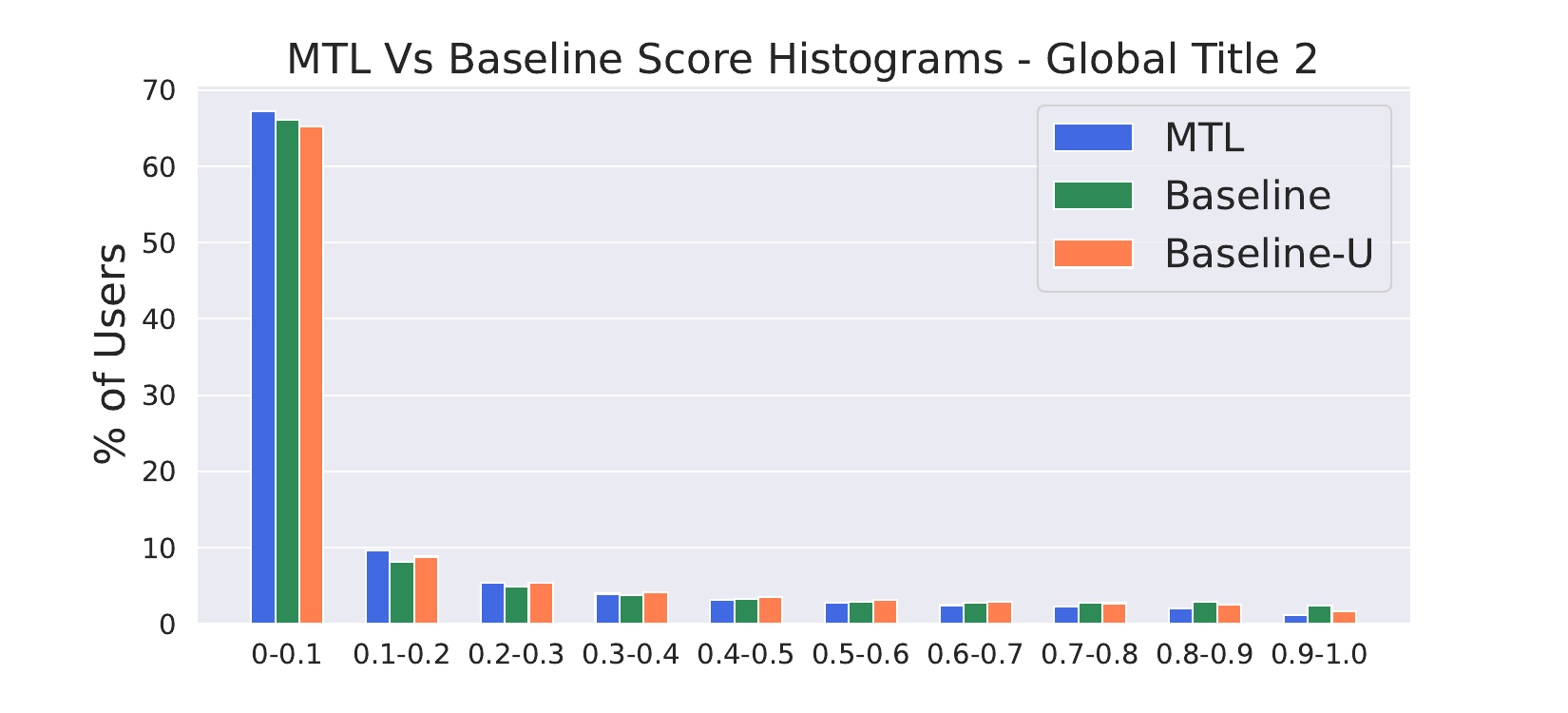}
    \caption{Global Title 2}
    \label{fig:br_global}
    \end{subfigure}
   
    \caption{Each subfigure shows the normalised user histograms of the model propensity scores of all the three models for a few titles. We can observe that the MTL-based model is giving relatively high scores for the local titles, and relatively low scores for the global titles attenuating their popularity bias.}
    \label{fig:score dist}
\end{figure*}

\section{Experiments}
\subsection{Data}
\label{subsec:experiments_data}

 For training and validation, we collected daily first time content discovery datasets for all the 200+ territories we are dealing with.~First time content discovery corresponds to generating positive user-title pairs based on the user’s first streaming event of title items within a month. We created two versions of the training dataset where several million users (order of 10M) are sampled randomly and then added the set of active users using the upsampling process described in section~\ref{subsec:Upsampling}. The training data consists of several hundred titles that were promoted
within the past several months spread across various territories. Overall, the user-title pairs for training are accumulated over several months time period. The number of negative samples is chosen to be $\alpha$ times bigger
than the number of positives, where $\alpha$ is approximately equal to the ratio of daily first time streamers among all Prime Video users. As given in table~\ref{table: data}, we can observe the data imbalance of positive rewards for all the titles in 16 major territories out of the total 200+. This in turn would result in disparity between titles that are present in multiple territories and titles that are present in few territories. We evaluate our models on three disjoint test data sets for three consecutive days post the end date of the training window. We sample a new set of users disjoint from the train in order of 10 millions for preparing the test data.

\subsection{Experiment Setup}
 For both the global and the MTL-based model, we use two dense layers along with batch-norm~\cite{ioffe2015batch} for each tower encoder. For task-specific layers in MTL, we use one dense layer for each task block. We used ReLU activation functions for all layers. We kept the hidden dimension as $d=512$ for all of the layers. For training, we use Adam~\cite{kingma2014adam} optimiser with a learning rate of 0.001 and a batch size of 8192. The models are trained end-to-end using the binary cross entropy loss as we are dealing with binary preferences. For MTL-based experiments, we use random batch strategy where each batch can contain examples from all the tasks. We use one-hot encoded mask while calculating the loss to ensure that only data points from the respective territory group can update the group-specific blocks, blocking the gradients of other blocks. We train the baseline model on both the normal and the upsampled data created using the approach described in section~\ref{subsec:experiments_data}. For MTL, we report the metrics for the model trained on upsampled data. To adapt the models to the daily patterns, we incrementally train all the models on ${i-1}^{th}$ day train data before evaluating on the $i^{th}$ day test data.
 \subsection{Results}
We compare the performance of MTL-based model against a baseline without MTL. As described in section~\ref{section:introduction}, our baseline model is a single global two-tower based model that finds the affinity score of a title given user using the dot product similarity. We chose PR-AUC metric for evaluation as it is more sensitive to the positive class performance, and is highly relevant when fraction of positives are much less than the negatives. As shown in the table~\ref{table: results}, MTL-based model consistently performs better than the single global model trained on both the normal and upsampled data in most of the territories. If we compare relative gains of MTL against the two baselines, we can observe that the baseline with upsampling improved over the baseline in a good number of territories. MTL performing better than the baseline trained on upsampled data demonstrates the advantages of the refined user representations that the MTL-based model learns.

\subsection{Case Study - Attenuating global popularity bias with improved local title affinity scores}
To further demonstrate the effectiveness of MTL, we analyze the user-title propensity score distributions for few selected global and local titles with comparable positive examples from two territories. We pick \textit{Local Title 1} and \textit{Global Title 1} from territory $T_i$ and \textit{Local Title 2} and \textit{Global Title 2} from territory $T_j$. 

First, we report the overall gain in title level PR-AUC score of the MTL-based model over both the baselines in table~\ref{table: case study}.
Then based on the normalised user histogram of output score bins given in figure~\ref{fig:score dist}, we try to answer these two questions: For local titles with comparable positive examples that of a global title, are we moving the score distributions towards the higher side? Yes, MTL-based model relatively moves more users from lower score bins to higher score bins as given in figures~\ref{fig:es_local} and~\ref{fig:br_local}. For global titles with comparable positive examples that of a local title, are we moving the score distributions towards the lower side? Yes, the MTL-based model relatively moves more number of users to the lower score bins from higher score bins attenuating the popularity bias as given in figures~\ref{fig:es_global} and~\ref{fig:br_global}.



\section{Conclusion}
In this paper, we propose an upsampling strategy based on active streamers strategy to deal with data imbalance. We also propose the use of an MTL-based architecture for learning refined user representations to better handle the distributional differences when dealing with multi-territory video recommendations. Through our experiments, we demonstrated the advantage of MTL-based architecture over a single global model trained on data from all the territories. Through our case study, we analysed the changes in the output score distributions of local and global titles thereby showing the advantages of the proposed techniques for attenuating the popularity bias. Overall, the proposed model achieves up to $65.27\%$ improvement in PR-AUC when evaluated on 16 major territories out of 200 overall territories.

	\bibliographystyle{plain}
	\bibliography{main.bib}
	
	\newpage

\end{document}